# GDNShift:
# Live Migration and Joint State Scheduling for Gated DeltaNet Inference

Hanqing Yu, Jiaqi Yang, Si Wu, Feng Li, Xiuzhen Cheng, Pengfei Hu, Yongbo Yu

## Abstract

Live migration of a streaming Gated DeltaNet (GDN) request requires rebuilding its recurrent state at the correct token boundary on a replica with identical weights. We propose a cross-replica recovery protocol based on compact update logs and implement and numerically validate exact catch-up. Compact logs stream a base state and ordered updates while the source continues decoding. Exact catch-up uses native single-token decode and releases the source once the destination has what it needs to continue independently. Across 120 complete pure-GDN configurations on three dual-GPU deployments, compact logs reduced the mean exact handoff gap from 790.8 to 149.5 ms relative to a background snapshot. Exact catch-up reduced mean source release time from 818.0 to 42.7 ms. In a prospective cross-host evaluation with relaxed arithmetic, a sealed selector chose the minimum-gap path in 39 of 40 configurations and reduced mean gap by 2.17% over a frozen network lookup. In nine whole-request migration paths for hybrid Qwen3.5-0.8B, the KV transfer ranking changed with the GDN path in 51 of 90 configurations. Per-configuration hindsight selection among the evaluated exact paths reduced mean gap by 34.7% over the best measured KV transfer paired with a GDN snapshot. In a separate held-out replay on 30 configurations, a locally calibrated selector reduced the exact gap by 15.4% over frozen network lookup.

## 1 Introduction

An inference server may need to go offline while it is still streaming a response. The destination must continue that response from the right position, and the operator needs to know when the source can be released. These events can occur at different times: once the destination has the inputs for independent continuation, the source can leave while the destination finishes recovery and computes the next token.

Ordinary attention appends key and value entries as the sequence grows. GDN updates a fixed-size recurrent state in place. Hybrid models also maintain short-convolution state and ordinary attention KV. All applicable state at handoff must correspond to the same consumed prefix. Existing attention-based systems migrate KV or send tokens for KV reconstruction. For GDN, a migration protocol must also determine how to convey the actively changing recurrent state and which numerical execution path will reconstruct it. [1] [2] [3] [4]

We study planned decode-to-decode migration between replicas with identical model weights. Our contributions are:

1. **GDN migration paths.** We propose a compact update-log protocol for cross-replica recovery. We implement exact catch-up and validate its numerical behavior. A background snapshot and 32-step batched replay provide two GDN adaptations for comparison.
2. **Pure-GDN trade-offs and prospective path selection.** We measure the paths under different objectives and numerical requirements, examine why compact logs and batched replay exchange ranks, and test a precommitted selector on the reverse cross-host deployment.

3 **Joint path selection for hybrid models.** We measure how the KV transfer method and GDN recovery path interact across nine complete request-migration paths. We quantify the value of a larger candidate set and evaluate probe-calibrated selection on held-out inputs.

# 2 Migration protocol

## 2.1 The handoff boundary

Let $P$ be the prefill length and $D$ the number of generated tokens visible when migration is requested. If $y_h$ is the final public token from source A, its model state has consumed the prompt and $y_1$ through $y_{h-1}$. Destination B restores that state and consumes $y_h$ to produce the first new public token, $y_{h+1}$. Intermediate replay outputs stay at B. The controller verifies coverage of state or update records, the final public token, and destination readiness.

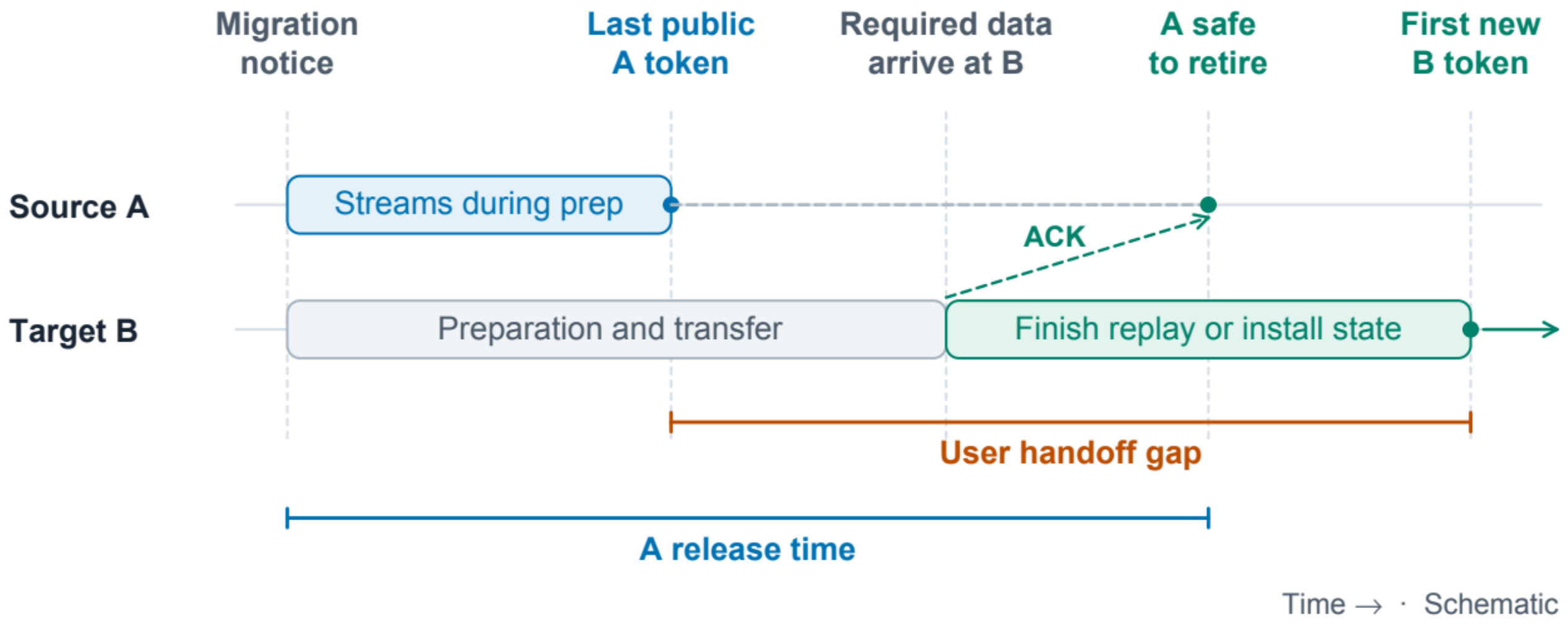

**Figure 1.** Migration notification, the last public A token, safe source release, and the first new B token are distinct events. The illustrated path can release A before B finishes computing.

## 2.2 Cross-replica recovery with compact update logs

A sends a base recurrent state and continues producing tokens. Each step sends the factors needed for its recurrent update. In the tested pure GDN model these include keys, values, decay, and the update gate. B applies the received records in order, executing the recurrent updates directly. The final message identifies the consumed boundary and carries the pending token. B checks update coverage and the pending token before taking over. Short-convolution state and ordinary attention KV, where present, align to the same prefix.

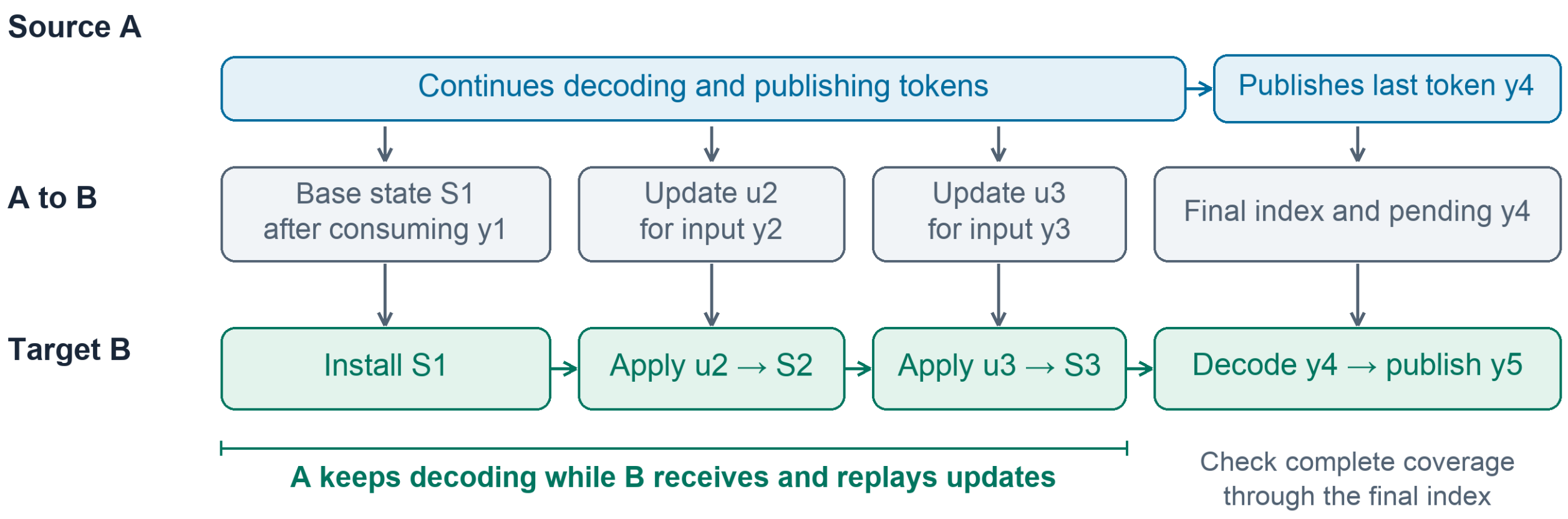

**Figure 2.** A streams a base and ordered updates while decoding. B can replay the updates as they arrive. Any updates still in flight or unapplied after A stops contribute to the handoff gap.

## 2.3 Exact catch-up implementation and numerical validation

At migration notification, the controller requests that A stop. For pure GDN, B needs the prompt, the tokens A has already published, and the final position. A KV-aligned hybrid path also confirms receipt of the ordinary attention KV it imports. B prefills the prompt and advances the same request through native single-token decode with A's public tokens. Once B has the data for independent continuation, A can be released. Any reconstruction that remains can extend the user's wait at B.

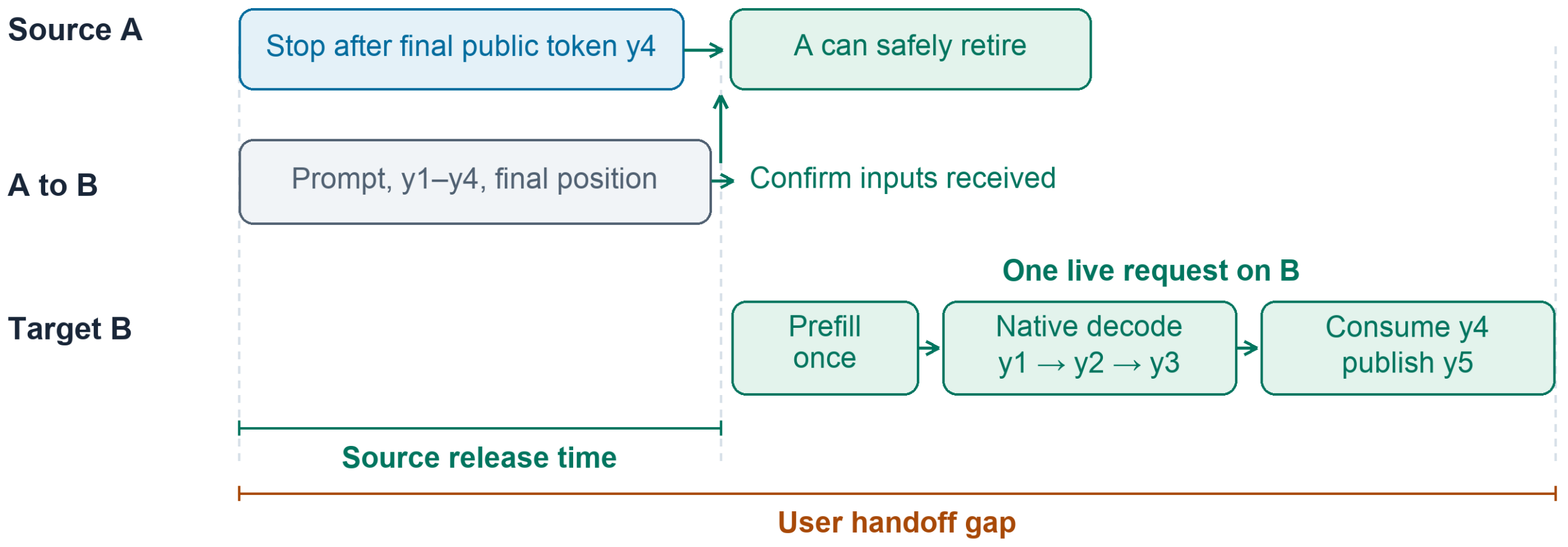

**Figure 3.** In the illustrated pure-GDN path, A can retire once B has the required inputs. B restores the consumed prefix, feeds the pending public token, and publishes the next new token.

Batched and single-token forwards can use different floating-point execution paths for the same tokens. In a diagnostic replay, identical normalized inputs reached the first-layer fused QKVZ projection. Of its 131,072 output elements, 92 differed, with a maximum absolute difference of 0.015625. In two supplementary tests with Qwen3.5-0.8B on an RTX 5060 Ti and SGLang with decode CUDA Graph enabled, A generated 80 tokens from a 2,048-token prompt and B continued for 256 steps. Recurrent state, short-convolution state, ordinary attention KV, full-vocabulary logits, and greedy output matched uninterrupted same-engine execution bitwise.

### 2.4 Adapted comparison paths

A **background snapshot** prepares B while A decodes, then transfers and installs the final recurrent state after A stops. **32-step batched replay** lets B recompute public history in chunks of at most 32 tokens, then replays the remaining segment through the final boundary. The snapshot preserves the transferred state at handoff. Batching reduces the number of model forwards and is eligible when arithmetic differences are allowed. The hybrid experiment adds a path that rebuilds both KV and GDN state in one destination prefill.

## 3 Pre-migration path selection

The selector takes the input lengths $P$ and $D$, link conditions, the numerical requirement, and the service objective. It filters paths by their correctness contract, then chooses for handoff gap, source release time, or the token rate during migration. Pure GDN uses four recurrent-state recovery paths. The hybrid candidate set contains complete KV-GDN combinations and the single-prefill rebuild path.

The pure-GDN estimator separates the wait for the last required packet from work at B after its arrival. For a packet of size $b_i$ enqueued at $a_i$, the FIFO model at application-layer rate $C$ gives the send completion time

$$e_i = \max(a_i, e_{i-1}) + b_i / C.$$

The estimator accounts for propagation and control separately and uses stage progress to predict how much work remains at the stop. When batched replay is expected to have 1-32 steps left, it uses the locally measured one-batch finishing time, including receiver-side wait, remaining computation, and first output. Crossing 32 steps may incur another batch. The reverse cross-host evaluation froze the rule before formal timing and substituted local probe parameters.

The hybrid selector defaults to a lookup table trained on the source deployment and indexed by bandwidth, round-trip time (RTT), objective, and numerical contract. It uses locally calibrated transfer, replay, and pre-copy estimates for exact gap or rate at 10 MB/s and for source release under relaxed arithmetic. Other conditions retain the lookup defaults. The target calibration uses 63 prior trials on three input shapes. The implementation uses the tested Qwen KV layout and the configured round budget.

## 4 Experimental design

### 4.1 Deployments and controls

| Experiment | Model and deployment | Measurement design |
|---|---|---|
| Pure-GDN main comparison | 1.3B; dual RTX 4090, cross-host RTX 3090 to RTX 4090, dual L40S | Each deployment has 9 input shapes × 5 links × (4 paths + no-migration control) × 3 repeats; 2,025 trial records |
| Prospective pure-GDN selection | 1.3B; reverse cross-host RTX 4090 to RTX 3090 | 40 held-out configurations, compact logs, batch replay, and no-migration control, three repeats each; 360 formal records |
| Pure-GDN checkpoint replication | 2B; dual RTX 4090 | Eight initial checks passed across two links; 440/675 timing records and 25 complete four-path configurations |
| Hybrid whole-request migration | Qwen3.5-0.8B; dual RTX 4090 and cross-host RTX 3090 to RTX 4090 | 9 inputs × 5 links × 9 paths × 3 repeats per deployment; 2,430 migration attempts across both, plus no-migration records for user-facing comparisons |

The total context length at notification, $P + D$, is 256, 1,024, or 1,536 with $P$:$D$ ratios of 7:1, 1:1, or 1:7. Link shaping configures 250 MB/s at 0.5 or 50 ms RTT, and 100, 50, or 10 MB/s at 50 ms RTT. The 1.3B checkpoint uses FLA 0.6.0 eager execution. The hybrid model uses SGLang. We take per-configuration medians of three valid repeats and weight configurations equally in the aggregate comparisons.

Of the 135 pure-GDN configurations in the three main deployments, exact catch-up reached a 30 s gap limit in 15. Joint rankings and means use the remaining 120 configurations with all four paths measured. A completed path can also be identified as the minimum-gap path when it beats a timeout lower bound and the other measured candidates. The hybrid per-configuration comparisons include paths with three complete measurements. The harness records interruption timeouts, source token capacity exhaustion, and pre-copy round-limit exits separately. The latter two establish feasibility limits under the configured implementation and budgets.

## 4.2 Metrics and acceptance checks

| Metric | Measurement |
|---|---|
| User handoff gap | Last public A token to first new B token |
| Source release time | Migration notification to confirmation that A can exit this request |
| Time to B handoff | Migration notification to first new B token, including continued generation at A |
| Token rate during migration | Count of new public tokens per unit time from the public token corresponding to notification to B's first token |

To estimate overhead against uninterrupted generation, we pair each migration with a no-migration trace of the same configuration and compare the time to produce the same number of new tokens. Different paths can cover different numbers of tokens by handoff. Comparing completion at a fixed output length would include the continuation on B.

We check the strict paths in a separate replay at the measured boundary. With matching GPUs, the reference is uninterrupted native decode in the same engine, with checks on applicable state, logits, and 48 free-running tokens. Across GPU architectures, snapshots and compact logs are checked against A's boundary state. Exact catch-up is checked against native replay of the same token trace on B. Relaxed paths still undergo boundary checks and checks for finite outputs. On selected acceptance inputs we terminate A's inference process at the safe-release acknowledgment and observe whether B continues independently.

# 5 Pure-GDN results

## 5.1 Method trade-offs

| Objective and numerical requirement | Winner in 120 complete configurations | Aggregate result |
|---|---|---|
| Minimum gap, strict | Compact logs, 120/120 | 790.8 to 149.5 ms vs. background snapshot, **81.1% lower** |
| Earliest source release, either requirement | Exact catch-up, 120/120 | 818.0 to 42.7 ms vs. background snapshot, **94.8% shorter** |
| Highest token rate, strict | Compact logs, 120/120 | 25.75 tokens/s vs. 37.57 without migration at matched token counts |
| Highest token rate, relaxed | 32-step batched replay, 120/120 | 33.43 tokens/s vs. 37.55 without migration at matched token counts |
| Minimum gap, relaxed | Compact logs 74/120, batched replay 46/120 | 69.5 ms mean with per-configuration hindsight selection |

Including the 15 configurations whose winner can be inferred from a timeout lower bound, compact logs achieve the minimum strict gap in 135/135. The source-release advantage of exact catch-up includes requesting the stop immediately. On same-host RTX 4090s at 100 MB/s and 50 ms RTT with $P = 128$, growing $D$ from 128 to 896 increases the exact catch-up gap from 3.52 to 24.48 s. Source release remains near 51 ms. The 25 complete 2B configurations reproduce the strict winner for minimum gap and the winner for early release.

## 5.2 Rank reversals and prospective selection

With relaxed arithmetic, link conditions change the gap winner. For dual RTX 4090, $P = D = 512$ and RTT fixed at 50 ms, compact logs and batched replay take 73.66 and 81.70 ms at 250 MB/s, and 443.90 and 83.84 ms at 10 MB/s. Approximately 246 ms of the slow-link difference comes from waiting for the last required packet and 114 ms comes from subsequent work and waiting at B. In a separate comparison, increasing the remaining replay from 32 to 33 steps raises B's finishing time from about 55-56 to 86.57 ms, with packet wait near 25.5 ms. Link backlog and batch boundaries can both change the choice.

On the reverse cross-host RTX 4090-to-3090 deployment, selection is restricted to compact logs and batched replay with relaxed arithmetic and minimum gap as the objective. We use the designated $P = D = 128$ geometry for calibration, seal predictions before formal timing, and score the remaining eight geometries on five links. The fixed path in the table is the strongest observed fixed choice across these configurations.

| Selection policy | Hits | Mean gap | Worst additional wait |
|---|---|---|---|
| Best fixed path, 32-step batched replay | 14/40 | 92.55 ms | 56.31 ms |
| Frozen network lookup | 36/40 | 80.16 ms | 23.98 ms |
| Frozen model with prior deployment parameters | 38/40 | 79.00 ms | 23.47 ms |
| Frozen model with local probes | **39/40** | **78.41 ms** | **0.68 ms** |
| Measured optimum by configuration | 40/40 | 78.40 ms | 0 ms |

Local probes save a mean 1.74 ms (2.17%) over network lookup. At 50 MB/s they save a mean 8.72 ms across eight inputs and improve hits from 5/8 to 8/8. For the other objectives on the complete pure-GDN matrix, a single fixed method per numerical requirement and objective already reaches the measured configuration-wise optimum.

# 6 Hybrid-model results

## 6.1 Value of joint path selection

We combine each of the four GDN recovery paths with either per-token KV transfer or multi-round KV pre-copy. A ninth path rebuilds both states in one prefill at B. The conventional baseline can choose the faster of the two KV transfers paired with a GDN snapshot for each configuration. Under relaxed arithmetic, it can also choose full rebuild. Comparing hindsight selection from the complete candidate set with this configuration-wise best baseline quantifies the opportunity provided by the additional GDN paths.

| Objective | Numerical requirement | Best baseline → best expanded candidate | Gain | Improved cells |
|---|---|---|---|---|
| Handoff gap | Path-specific bitwise validation | 669.6 → 437.4 ms | **34.7% lower** | 80/90 |
| Handoff gap | Relaxed | 52.69 → 49.42 ms | **6.2% lower** | 49/90 |
| Source release time | Path-specific bitwise validation | 1,034.5 → 370.9 ms | **64.1% shorter** | 90/90 |
| Source release time | Relaxed | 85.22 → 81.37 ms | **4.5% shorter** | 12/90 |
| Token rate during migration | Path-specific bitwise validation | 81.98 → 123.26 tokens/s | **50.4% higher** | 78/90 |
| Token rate during migration | Relaxed | 114.51 → 176.75 tokens/s | **54.4% higher** | 90/90 |

Among strict-path gap comparisons, 51 of 90 configurations exhibit a KV ranking reversal across GDN paths. We count a reversal when, for each of two GDN paths, the two KV methods have non-overlapping ranges over three repeats and opposite orderings. For RTX 3090 to RTX 4090 at $P = D = 768$, 100 MB/s, and 50 ms RTT, the snapshot gap is 285.44 ms with pre-copy and 302.61 ms with per-token KV. With compact logs, per-token KV takes 55.15 ms and pre-copy takes 83.63 ms. Taking the KV winner measured with snapshots and pairing it with compact logs therefore adds 28.48 ms over the best measured complete path in that configuration.

Packet records show that large KV packets can queue ahead of log updates on the shared link. Destination progress records also show that the arrival schedule of KV can govern how much GDN replay B completes before the cut. These observations support queuing and readiness effects. A controlled experiment is needed to isolate GPU contention.

## 6.2 Held-out replay with local calibration

We determine the formula and lookup structure from RTX 3090-to-4090 data and recalibrate parameters on three inputs on dual RTX 4090. The held-out evaluation covers six other input shapes and five links. It scores six decisions on each of the same 30 configurations using the measured outcomes of the chosen paths. The table excludes calibration cost from the path timing.

| Objective | Numerical requirement | Lookup hits → formula hits | Measured means |
|---|---|---|---|
| Handoff gap | Path-specific bitwise validation | 28/30 → **30/30** | 546.98 → 463.02 ms |

| Objective | Numerical requirement | Lookup hits → formula hits | Measured means |
|---|---|---|---|
| Handoff gap | Relaxed | 26/30 → 26/30 | 49.11 → 49.11 ms |
| Source release time | Path-specific bitwise validation | 29/30 → 29/30 | 372.21 → 372.21 ms |
| Source release time | Relaxed | 26/30 → **30/30** | 83.74 → 80.11 ms |
| Token rate during migration | Path-specific bitwise validation | 23/30 → **25/30** | 132.44 → 134.97 tokens/s |
| Token rate during migration | Relaxed | 20/30 → 20/30 | 187.62 → 187.62 tokens/s |

Aggregate hits rise from 152/180 to 160/180. The mean strict-path gap is 15.4% shorter using unrounded means. This reduction comes mainly from two short-input configurations at 10 MB/s, with savings near 0.90 and 1.62 s. Objectives that choose the same default path retain the same measured performance. Prospective end-to-end execution and net gains after probe costs remain to be evaluated.

### 6.3 Timing and user experience

Among strict paths, minimum user gap and earliest first output at B select different paths in 73/90 configurations. The per-configuration fastest token-rate paths average 1.75 s until B's first output, cover about 169 tokens, and deliver 123.2 tokens/s. The fixed per-token KV plus GDN snapshot baseline averages 1.00 s to B's first output, spans about 68 tokens, and delivers 61.4 tokens/s. Continued source generation changes how many tokens are visible before handoff. Token rate in the migration window and time to destination handoff are distinct service objectives.

## 7 Related work and contribution boundary

Llumnix migrates live attention-based requests through KV transfer. ServerlessLLM migrates intermediate tokens in multiple rounds and recomputes KV at the destination. Its protocol distinguishes unloading the source model from resuming at the destination. Marconi studies prefix reuse across hybrid-model requests and identifies the constraints of in-place recurrent state updates. Our setting is live decode-to-decode migration of an active GDN request between same-weight replicas. [2] [3] [4]

In the SGLang community, ReplaySSM already uses a GDN checkpoint and compact buffer within one worker for ordinary decode and uses ordered recurrence replay in speculative decoding. A Qwen3-8B attention-only prototype uses native single-token replay when handing off from prefill to decode, with the prefill worker able to stream while prompt KV is in flight. Another community PR transfers GDN state at a prefill-to-decode handoff. Compact logs contribute the cross-replica recovery protocol for streaming recurrent updates and verifying their coverage at handoff. Exact catch-up contributes the GDN system implementation and its numerical validation, including the applicable references and continuation after source release. [C1] [C2] [C3] [C4]

# 8 Limitations and follow-up experiments

The experiments cover planned single-request migration. The pure-GDN matrix uses three deployments, supplemented by the reverse cross-host selector evaluation and partial 2B checkpoint replication. The nine-path hybrid evaluation uses one checkpoint. Three repetitions yield per-configuration medians. Near ties, particularly around the 50-100 MB/s transition, warrant more interleaved repetitions. Link rates are software configurations. The reported rankings include the behavior of the transport and recovery paths under those settings. Hybrid candidates also depend on the current KV layout, capacity, and pre-copy round budget.

1. **Prospective hybrid evaluation.** Freeze the selector, candidate set, calibration budget, and comparisons on a new deployment, then run the selected paths. Compare frozen lookup, a simple rule given the same local calibration budget, and the strongest fixed pair selected from development data. Measure calibration, decision time, failure rates, and the observed gaps separately.
2. **Isolating the KV-GDN interaction.** Fix the handoff position and amount of work on representative reversals. Vary KV availability, packet ordering, or destination compute independently and measure the effects on last data arrival and first output. Sweep representative pre-copy settings.
3. **Checkpoint and workload coverage.** Compare additional pure-GDN checkpoints on the same hardware and hybrid models with different KV sizes and layer mixes. For multi-model serving, also measure impact on requests that remain on A and resource release on the source. Fixed-length completion and selection under a gap bound are additional evaluation targets.

# 9 Conclusion

Compact update logs and exact catch-up improve different parts of live GDN migration on the tested deployments: the exact handoff gap and source release time. Under relaxed arithmetic, either logs or batched replay can deliver the shorter pure-GDN gap. Locally calibrated prospective selection improves decisions in the competing link regime. In hybrid models, KV transfer and GDN recovery interact, as shown by measured ranking reversals across complete paths. Held-out replay shows measurable gains from local calibration for a subset of objectives.

## Open-source community implementations